\documentclass[twocolumn,nofootinbib,amsmath,amssymb,a4paper]{revtex4}

\usepackage{graphicx}
\usepackage{dcolumn}
\usepackage{bm}

\begin{document}

\title{
Study of Penguin Pollution in the $B^0\to J/\psi K_S$
Decay~\footnote{Talk presented at the 4th International Workshop on the
CKM Unitarity Triangle (CKM2006), Nagoya, Japan, Dec.~12-16, 2006, which
is based on the work in Ref~\cite{Li:2006vq}, collaborated with
H-n.~Li.} 
}

\author{Satoshi Mishima}
\email{mishima@ias.edu}

\affiliation{%
School of Natural Sciences, Institute for Advanced Study, 
Princeton, NJ 08540, U.S.A.
}%

\begin{abstract}
We study the penguin pollution in the $B^0\to J/\psi K_S$ decay up to
 leading power in $1/m_b$ and to next-to-leading order in $\alpha_s$,
 $m_b$ being the $b$ quark mass and $\alpha_s$ the strong coupling
 constant. The deviation $\Delta S_{J/\psi K_S}$ of the mixing-induced
 CP asymmetry from $\sin(2\phi_1)$ and the direct CP asymmetry
 $A_{J/\psi K_S}$ are both found to be of $O(10^{-3})$ in a formalism
 that combines the QCD-improved factorization and perturbative QCD
 approaches. 
\end{abstract}

\maketitle

\section{Introduction}

The $B^0\to J/\psi K_S$ decay is known to be the golden mode for
extracting $\sin(2\phi_1)$, $\phi_1$ being the weak phase of the
Cabbibo-Kobayashi-Maskawa (CKM) matrix element $V_{td}$, through the
time-dependent CP asymmetry  
\begin{eqnarray}
a(t)
 &=&
\frac{\Gamma({\bar B}^0(t)\to {J/\psi K_S})- \Gamma(B^0(t)\to {J/\psi K_S})}
{\Gamma({\bar B}^0(t)\to {J/\psi K_S})+\Gamma(B^0(t)\to {J/\psi K_S})} 
\;,\nonumber\\
&=&
S_{J/\psi K_S}\sin(\Delta M t)
+ A_{J/\psi K_S}\cos(\Delta M t)
\;, 
\label{eq:asymmetry}
\end{eqnarray}
where $\Delta M$ is the mass difference of the two $B_d$-meson mass
eigenstates. The mixing-induced CP asymmetry $S_{J/\psi K_S}$ is
na\"{\i}vely equal to $\sin(2\phi_1)$ in the standard model. At the same
time, the direct CP asymmetry $A_{J/\psi K_S}$ is expected to be
vanishingly small. The penguin pollution, which would change the above
na\"{\i}ve expectations, is believed to be negligible in this mode. A
complete estimation of its effect is, however, essential for future
precision measurements, where the experimental error of $S_{J/\psi K_S}$
will be $\delta S_{J/\psi K_S}\approx \pm 0.2$ for 5\,ab$^{-1}$ data at
the Super $B$ factory~\cite{Akeroyd:2004mj}.

The deviation $\Delta S_{J/\psi K_S}\equiv S_{J/\psi K_S}-\sin(2\phi_1)$
was found to be of $O(10^{-4})$ in the previous estimation by Boos 
{\it et al.}~\cite{Boos:2004xp}, taking into account corrections to the
$B-\bar B$ mixing and to the $B^0\to J/\psi K^0$ decay amplitude. A part
of penguin-operator contributions, however, were overlooked in the
estimation.

A model-independent approach to estimate the size of the penguin
pollution proposed by Ciuchini {\it et al.}~\cite{Ciuchini:2005mg}. They
extracted the range of the penguin corrections from the data of the
branching ratio and the CP asymmetries for the $B^0\to J/\psi\pi^0$
decay, and used the range to evaluate $\Delta S_{J/\psi K_S}$ with the
flavor SU(3) symmetry. The current data lead to 
\begin{equation}
\Delta S_{J/\psi K_S}^{\rm decay} = 0.000\pm 0.012
\;, 
\end{equation}
where the large error comes from the experimental uncertainty of the
$B^0\to J/\psi\pi^0$ data.

In this talk, we present the most complete analysis of the penguin
pollution in $B^0\to J/\psi K_S$ up to leading power in $1/m_b$ and to
next-to-leading order (NLO) in $\alpha_s$, $m_b$ being the $b$ quark
mass and $\alpha_s$ the strong coupling constant. We shall concentrate
only on the corrections to the $B-\bar B$ mixing and to the decay
amplitude, and refer the inclusion of corrections to the $K - \bar K$
mixing and to the $B$-decay-width difference $\Delta\Gamma_B$ to
Ref.~\cite{Grossman:2002bu}.

This talk is organized as follows: 
In Sec.~\ref{sec:mixing}, we briefly review the corrections to the
$B-\bar B$ mixing. The corrections to the decay amplitude are studied in
Sec.~\ref{sec:decay}. Numerical results are then presented in
Sec.~\ref{sec:results}. 
Section~\ref{sec:conclusion} is the conclusion.

\section{Corrections to the $B-\bar B$ Mixing \label{sec:mixing}}

\begin{figure}
\includegraphics[width=0.20\textwidth]{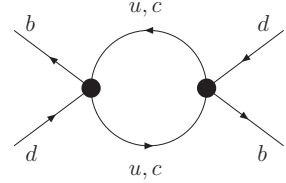}
\caption{Non-local corrections to the $B-\bar B$ mixing, where the dots
 represent local operators. \label{fig:mixing}}
\end{figure}

The mixing-induced CP asymmetry $S_{J/\psi K_S}$ and the direct CP
asymmetry $A_{J/\psi K_S}$ in Eq.~(\ref{eq:asymmetry}) are given by 
\begin{eqnarray}
S_{J/\psi K_S}=
\frac{2\,{\rm Im}\,\lambda_{J/\psi K_S}}
{1+|\lambda_{J/\psi K_S}|^2}\;,
\nonumber\\
A_{J/\psi K_S}=
\frac{|\lambda_{J/\psi K_S}|^2-1}
{1+|\lambda_{J/\psi K_S}|^2}\;,
\end{eqnarray}
with the associated factor 
\begin{equation}
\lambda_{J/\psi K_S} = 
-\frac{q}{p}\, 
\frac{{\cal A}({\bar B}^0 \to J/\psi K_S)}
{{\cal A}(B^0 \to J/\psi K_S)}\;,
\end{equation}
where the ratio $q/p$ is related to the $B-\bar B$ mixing and 
${\cal A}(B^0({\bar B}^0) \to J/\psi K_S)$ is the decay amplitude. 
The corrections to the $B-\bar B$ mixing alters $q/p$ from
$\exp(-2i\phi_1)$. In Ref.~\cite{Boos:2004xp}, non-local contributions
to the $B-\bar B$ mixing, shown in Fig.~\ref{fig:mixing}, were calculated. 
The non-local operators are defined as  
\begin{eqnarray}
T_1 = 
- \frac{i}{2}
\int d^4x\,{\rm T}\left[
[(\bar b\,c)_\mu(\bar c\,d)^\mu](x)
[(\bar b\,c)_\nu(\bar c\,d)^\nu](0)
\right],
\nonumber\\
T_2 = 
- \frac{i}{2}
\int d^4x\,{\rm T}\left[
[(\bar b\,c)_\mu(\bar c\,d)^\mu](x)
[(\bar b\,u)_\nu(\bar u\,d)^\nu](0)
\right],
\nonumber\\
T_3 = 
- \frac{i}{2}
\int d^4x\,{\rm T}\left[
[(\bar b\,u)_\mu(\bar u\,d)^\mu](x)
[(\bar b\,u)_\nu(\bar u\,d)^\nu](0)
\right],
\end{eqnarray}
where the left-handed current is shortened into 
$(\bar b\,q)_\mu \equiv (\bar b_L\gamma_\mu q_L)$, and ${\rm T}$
denotes the time-ordered product. 
As the scale evolves down, the non-local operators mix into the 
$\Delta B=2$ local operators of dimension 8, 
\begin{eqnarray}
Q_1&=&\Box 
(\bar b\, d)_\mu (\bar b\, d)^\mu
\;,\nonumber\\
Q_2&=&\partial^\mu\partial^\nu
(\bar b\, d)_\mu (\bar b\, d)_\nu
\;,\nonumber\\
Q_3&=& m_c^2
(\bar b\, d)_\mu (\bar b\, d)^\mu
\;,
\end{eqnarray}
$m_c$ being the $c$-quark mass, through renormalization-group
effects. Including QCD corrections at $O(\alpha_s)$ and resumming the
large logarithm $\alpha_s\ln(m_W/m_b)$, where $m_W$ is the $W$-boson mass,
the corrections to the mixing factor $q/p$ lead to~\cite{Boos:2004xp} 
\begin{eqnarray}
\Delta S_{J/\psi K_S}^{\rm mix}
&=& \left(2.08\pm 1.23 \right)\times 10^{-4}
\;,
\nonumber\\
A_{J/\psi K_S}^{\rm mix} 
&=& \left(2.59 \pm 1.48 \right)\times 10^{-4}
\;. 
\label{eq:results-mixing}
\end{eqnarray}
Thus, $\Delta S_{J/\psi K_S}^{\rm mix}$ and $A_{J/\psi K_S}^{\rm mix}$
have been both found to be of $O(10^{-4})$.

\section{Corrections to the Decay Amplitude \label{sec:decay}}

The $B^0 \to J/\psi K^0$ decay amplitude can be decomposed into 
\begin{eqnarray}
&&
{\cal A}(B^0 \to J/\psi K^0)
\nonumber\\
&&=
V_{ub}^*V_{us}\, {\cal A}^{(u)}_{J/\psi K^0}
+V_{cb}^*V_{cs}\, {\cal A}^{(c)}_{J/\psi K^0}
+V_{tb}^*V_{ts}\, {\cal A}^{(t)}_{J/\psi K^0}
\;,\nonumber\\
&&=
\sum_{q=u,c}
V_{qb}^*V_{qs}\, 
\left( 
 {\cal A}^{(q)}_{J/\psi K^0} - {\cal A}^{(t)}_{J/\psi K^0} 
\right), 
\label{eq:amp}
\end{eqnarray} 
where the unitarity relation of the CKM matrix elements 
$V_{tb}^*V_{ts}=-V_{ub}^*V_{us}-V_{cb}^*V_{cs}$ is used. 
The term with $V_{cb}^*V_{cs}$ dominates the decay amplitude, whereas
that with $V_{ub}^*V_{us}$ causes the penguin pollution, since it
carries the CKM phase $\phi_3$, which is defined by
$V_{ub}=|V_{ub}|\exp(-i\phi_3)$. Note that the amplitude
$V_{ub}^*V_{us}{\cal A}^{(t)}_{J/\psi K^0}$ was missed in 
Ref.~\cite{Boos:2004xp}.

\begin{figure}
\includegraphics{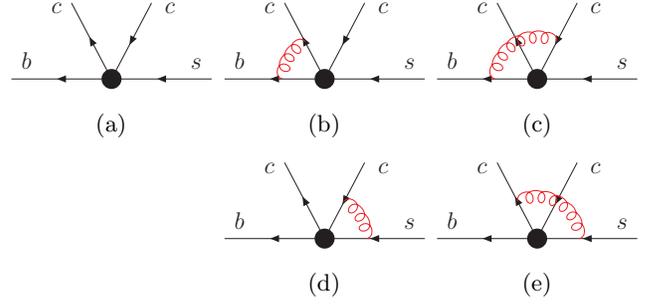}
\caption{Factorizable diagrams, where the spectator quark is
 omitted. \label{fig:fact}} 
\end{figure}
We calculate the decay amplitude using a formalism~\cite{Chen:2005ht}  
that combines the QCD-improved factorization (QCDF)~\cite{BBNS} and the
perturbative QCD approach (PQCD)~\cite{PQCD} up to leading power in
$1/m_b$ and to NLO in $\alpha_s$. 
We adopt QCDF to handle the factorizable amplitudes, since the energy
release in the form factor $F_+^{BK}(m_{J/\psi}^2)$, $m_{J/\psi}$ being
the $J/\psi$ meson mass, is small and it is unlikely to further
factorize the form factor. 
The factorizable contributions to ${\cal A}^{(c)}_{J/\psi K^0}$ and
${\cal A}^{(t)}_{J/\psi K^0}$ displayed in Fig.~\ref{fig:fact} are given
by   
\begin{eqnarray}
{\cal A}^{(c)f}_{J/\psi K^0}&=& 
2\sqrt{6}\int_0^1 dx_c
\Psi^L(x_c)F_+^{BK}(m_{J/\psi}^2)
a_2(x_c,t)
\;,\nonumber\\
{\cal A}^{(t)f}_{J/\psi K^0}&=&
2\sqrt{6}\int_0^1 dx_c\,
\Psi^L(x_c)
F_+^{BK}(m_{J/\psi}^2)
\nonumber\\
&&\ \ \ \ \ \ \ \ \ \ \ \ \ \ 
\times \left[a_3(x_c,t)+a_5(x_c,t)\right], 
\label{eq:fact}
\end{eqnarray}
respectively, where $x_c$ is the momentum fraction carried by the $c$
quark, and $\Psi^L(x_c)$ the twist-2 $J/\psi$ meson distribution
amplitude. The explicit expressions of $\Psi^L(x_c)$ is referred to
Ref.~\cite{Li:2006vq}.  
The hard scale $t$ has been chosen as 
\begin{eqnarray}
t = \max\left(\sqrt{x_c(m_B^2-m_{J/\psi}^2)},
\sqrt{\overline{x_c}(m_B^2-m_{J/\psi}^2)}\ \right), 
\end{eqnarray}
with the notation $\overline{x_c}=1-x_c$ and $m_B$ being the $B$ meson
mass.  
The effective Wilson coefficients, including the $O(\alpha_s)$ vertex
corrections in Figs.~\ref{fig:fact}(b)-(d), are defined as 
\begin{eqnarray}
a_2(x,\mu)&=&
C_1(\mu)+\frac{C_2(\mu)}{N_c}
\nonumber\\
&&\hspace{-10mm} 
\times \left[1+\frac{\alpha_s(\mu)}{4\pi}C_F
\left(-18+12\ln\frac{m_b}{\mu}+ f_I(x)\right)\right],
\nonumber\\
a_3(x,\mu)&=&
C_3(\mu)+C_9(\mu)+\frac{1}{N_c}\left[C_4(\mu)+C_{10}(\mu)\right]
\nonumber\\
&&\hspace{-10mm} 
\times \left[1+\frac{\alpha_s(\mu)}{4\pi}C_F
\left(-18+12\ln\frac{m_b}{\mu}+f_I(x)\right)\right],
\nonumber\\
a_5(x,\mu)&=&
C_5(\mu)+C_7(\mu)+\frac{1}{N_c}\left[C_6(\mu)+C_8(\mu)\right]
\nonumber\\
&&\hspace{-10mm} 
\times \left[1+\frac{\alpha_s(\mu)}{4\pi}C_F
\left(6-12\ln\frac{m_b}{\mu}-f_I(x)\right)\right],
\end{eqnarray}
with $N_c=3$ being the number of colors, $C_F=4/3$ the color factor, 
and $C_{1-10}$ the Wilson coefficients defined, {\it e.g.}, in
Ref.~\cite{PQCD}. The loop function $f_I(x)$ is given by~\cite{JPSIK} 
\begin{eqnarray}
f_I(x)
&=&\frac{3(1-2x)}{1-x}\ln x-3\pi i
\nonumber\\
&&\ \ \ \ \ \ \ \
+3\ln(1-r_{J/\psi}^2)+\frac{2r_{J/\psi}^2(1-x)}{1-r_{J/\psi}^2 x}
\;,
\end{eqnarray}
with the ratio $r_{J/\psi}=m_{J/\psi}/m_B$.

\begin{figure}
\includegraphics{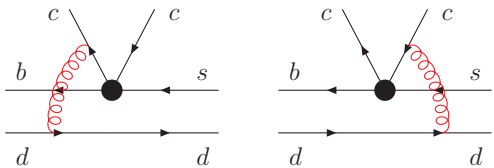}
\caption{Nonfactorizable spectator diagrams. \label{fig:nonfact}}
\end{figure}
For the $O(\alpha_s)$ nonfactorizable spectator diagrams in 
Fig.~\ref{fig:nonfact}, QCDF is not appropriate due to the end-point
singularity from vanishing parton momenta. 
Note that the nonfactorizable contribution has a characteristic scale
higher than in $F_+^{BK}(m_{J/\psi}^2)$~\cite{Chou:2001bn}. 
Therefore, we can employ PQCD based on $k_T$ factorization theorem,
which is free of the end-point singularity.  
The nonfactorizable spectator amplitudes in PQCD are given by 
\begin{eqnarray}
{\cal A}^{(c)nf}_{J/\psi K^0}&=&
{\cal M}\left(a_2^\prime \right),
\nonumber\\
{\cal A}^{(t)nf}_{J/\psi K^0}&=&
{\cal M}\left(a_3^\prime-a_5^\prime \right),
\label{eq:nonfact}
\end{eqnarray}
where the explicit expression of the amplitude ${\cal M}$ is referred to 
Refs.~\cite{Li:2006vq,Chen:2005ht}, and the effective Wilson
coefficients are defined by 
\begin{eqnarray}
a_{2}^{\prime }(\mu) 
&=&
\frac{C_{2}(\mu)}{N_{c}}
\;,\nonumber\\
a_{3}^{\prime }(\mu) 
&=&
\frac{1}{N_{c}} 
\left[ C_{4}(\mu)+C_{10}(\mu)\right]
,\nonumber\\
a_{5}^{\prime}(\mu)
&=&
\frac{1}{N_{c}}\left[ C_{6}(\mu)+C_{8}(\mu)\right]
.
\end{eqnarray}
The nonfactorizable contribution is essential for the 
$B^0\to J/\psi K^0$ decay, since this mode is classified in the
color-suppressed category of $B$ meson decays. 
In fact, the nonfactorizable contribution is comparable in size to the
factorizable one.

\begin{figure}
\includegraphics{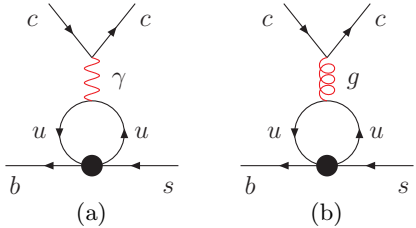}
\caption{$u$-quark loop diagrams (a) with a photon emission and (b) with
 a gluon emission. \label{fig:upenguin}}
\end{figure}

The amplitude ${\cal A}^{(u)}_{J/\psi K^0}$ receives the contribution
from the $u$-quarks loop shown in Figs.~\ref{fig:upenguin}(a) and (b). 
The penguin corrections from the $u$-quark loops as well as from the 
$c$-quark ones are well-behaved in perturbation theory without any
infrared singularity, which has been known as the Bander-Silverman-Soni
mechanism~\cite{Bander:1979px}. It means that their long-distance
contribution is unlikely to be large. The light-cone-sum-rules analysis
has also suggested that these corrections are dominated by short-distance
contribution~\cite{Khodjamirian:2003eq}. The $u$-quark loop contribution
was estimated na\"{\i}vely as  
$\Delta S_{J/\psi K_S}^{\rm decay}= - (4.24\pm 1.94)\times
10^{-4}$~\cite{Boos:2004xp}. We shall reinvestigate this contribution
below.  

\begin{figure}
\includegraphics{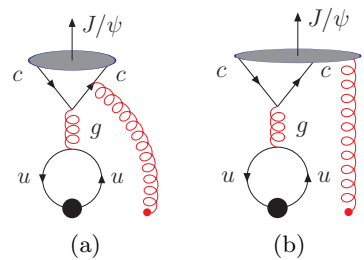}
\caption{QCD $u$-quark loop diagrams (a) with an additional hard gluon
 and (b) with a soft gluon. \label{fig:upenguin-qcd}}
\end{figure}

The $u$-quark loop contribution can be expressed in terms of the
effective Hamiltonian 
\begin{eqnarray}
&&{\cal H}_{\rm eff}^{(u)}\, =\, 
-\frac{G_F}{\sqrt{2}}V_{us}^*V_{ub}
\left(\frac{2}{3}-\ln\frac{l^2}{\mu^2}+i\pi\right)
\nonumber\\ 
&&\hspace{3mm}
\times 
\bigg[
\frac{\alpha}{3\pi}e_ue_c \left(N_cC_1(\mu)+C_2(\mu)\right)
\left(\bar c \gamma_\mu c \right)
\left(\bar s \gamma^\mu(1-\gamma_5) b \right)
\nonumber\\ 
&&\hspace{8mm}
+\, 
\frac{\alpha_s}{3\pi}C_2(\mu)
\left(\bar c \gamma_\mu T^a c \right)
\left(\bar s \gamma^\mu(1-\gamma_5) T^a b \right)
\bigg]\;,
\label{eq:Heff}
\end{eqnarray}
where $G_F$ is the Fermi constant, $\alpha$ the fine-structure constant,
$e_{u(c)}=2/3$ the electric charge of the $u(c)$ quark, and $T^a$ the
SU(3) generator. 
The first and second terms arise from
the photon emission diagram in Fig.~\ref{fig:upenguin}(a) and the gluon
emission one in Fig.~\ref{fig:upenguin} (b), respectively. $l^2$ denotes
the invariant mass squared of the photon or gluon. The photon emission
contribution is less than 5\% of ${\cal A}^{(t)}_{J/\psi K}$ due to the
smallness of $\alpha$, and can be safely dropped.  
For the gluon emission, an additional gluon is necessary to form the
color-singlet $J/\psi$ meson. If the additional gluon is hard, the
resultant contribution in Fig.~\ref{fig:upenguin-qcd}(a) is of
next-to-next-to-leading order in $\alpha_s$ and beyond the scope of this
work. If the additional gluon is soft, the corresponding nonperturbative
input is the three-parton $J/\psi$ meson distribution amplitude as
shown in Fig.~\ref{fig:upenguin-qcd}(b). The twist-3 distribution
amplitude, which contributes only to transversely polarized $J/\psi$
mesons, is irrelevant here. The twist-4 distribution amplitudes are
antisymmetric under the exchange of the momentum fractions of the $c$
quark and of the $\bar c$ quark. Because the hard kernel associated with
the $u$-quark loop is symmetric under the above exchange, its
convolution with the distribution amplitudes diminishes. Therefore, the
gluon emission contribution as well as the photon emission one can be
neglected.  
The $c$-quark loop corrections can also be neglected, since it modifies
only the branching ratio slightly.

In summary, the amplitudes in Eq.~(\ref{eq:amp}) are given by 
\begin{eqnarray}
{\cal A}^{(u)}_{J/\psi K^0}
&\simeq& 
0
\;,\nonumber\\
{\cal A}^{(c)}_{J/\psi K^0}
&\simeq& 
{\cal A}^{(c)f}_{J/\psi K^0} + {\cal A}^{(c)nf}_{J/\psi K^0}
\;,\nonumber\\
{\cal A}^{(t)f}_{J/\psi K^0}
&\simeq& 
{\cal A}^{(t)f}_{J/\psi K^0} + {\cal A}^{(t)nf}_{J/\psi K^0}
\;,
\end{eqnarray}
where ${\cal A}^{(c,t)f}$ and ${\cal A}^{(c,t)nf}$ are found in
Eqs.~(\ref{eq:fact}) and (\ref{eq:nonfact}), respectively.

\section{Numerical Results \label{sec:results}}

We predict the branching ratio, $\Delta S_{J/\psi K_S}$, and 
$A_{J/\psi K_S}$, with the corrections to the decay amplitude, 
\begin{eqnarray}
\hspace{-4mm}
{\rm Br}(B^0\to J/\psi K^0) &=&
\left( 6.6^{+3.7\,(+3.7)}_{-2.3\,(-2.3)} \right)\times 10^{-4},
\nonumber\\
\Delta S_{J/\psi K_S}^{\rm decay} &=&
\left( 7.2^{+2.4\,(+1.2)}_{-3.4\,(-1.1)} \right)\times 10^{-4},
\nonumber\\
A_{J/\psi K_S}^{\rm decay} &=&
-\left( 16.7^{+6.6\,(+3.8)}_{-8.7\,(-4.1)} \right)\times 10^{-4},
\label{eq:results-decay}
\end{eqnarray}
where the choices of parameters used in the computation, including their
allowed ranges, are referred to Ref.~\cite{Li:2006vq}. 
The errors are estimated from the variation of the CKM and hadronic
parameters, while those in the parentheses arise only from the hadronic
ones.  
The predicted branching ratio is in agreement with the data 
${\rm Br}(B^0\to J/\psi K^0) = (8.63\pm 0.35)\times 10^{-4}$~\cite{HFAG}. 
The form factor $F^{BK}_{+}(m^2_{J/\psi})$ increased by 15\% and larger
Gegenbauer moments in the kaon distribution amplitudes can easily
account for the central value of the data. 
Our $\Delta S_{J/\psi K_S}^{\rm decay}$ ($A_{J/\psi K_S}^{\rm decay}$)
from the penguin corrections is about twice of the na\"{\i}ve estimation
from the $u$-quark loops~\cite{Boos:2004xp}, and has an opposite (the
same) sign. Both $\Delta S_{J/\psi K_S}^{\rm decay}$ and 
$A_{J/\psi K_S}^{\rm decay}$ in Eq.~(\ref{eq:results-decay}) indicate
the $O(10^{-3})$ penguin pollution in the $B^0\to J/\psi K_S$ decay,
consistent with the conjecture made in Ref.~\cite{Grossman:2002bu}.

Including the correction to the $B-\bar B$ mixing in
Eq.~(\ref{eq:results-mixing}), we predict $\Delta S_{J/\psi K_S}$ and
$A_{J/\psi K_S}$, up to leading-power in $1/m_b$ and to NLO in
$\alpha_s$, as 
\begin{eqnarray}
\Delta S_{J/\psi K_S}
&=&
\Delta S_{J/\psi K_S}^{\rm mix}
+\Delta S_{J/\psi K_S}^{\rm decay}
\;,\nonumber\\
&=&
\left( 9.3^{+3.6}_{-4.6} \right) \times 10^{-4},
\nonumber\\
A_{J/\psi K_S}
&=&
A_{J/\psi K_S}^{\rm mix}+A_{J/\psi K_S}^{\rm decay}
\;,\nonumber\\
&=&
-\left(14.1^{+8.1}_{-10.2} \right) \times 10^{-4}. 
\end{eqnarray}
Taking into account the CP violation from the $K-\bar K$
mixing~\cite{Grossman:2002bu}, $\Delta S_{J/\psi K_S}$ and 
$A_{J/\psi K_S}$ remain $O(10^{-3})$.

\section{Conclusion \label{sec:conclusion}}

In this talk, we have presented the most complete analysis of the
branching ratio and the CP asymmetries for the $B^0\to J/\psi K_S$ decay 
up to leading power in $1/m_b$ and to NLO in $\alpha_s$. 
We have found that both $\Delta S_{J/\psi K_S}$ and $A_{J/\psi K_S}$ are
of $O(10^{-3})$.   
Our prediction for $\Delta S_{J/\psi K_S}$ is smaller than an expected
systematic error in future data~\cite{NAKAMURA}.  
As for the direct CP asymmetry, our result supports the claim that
$A_{J/\psi K_S}\,\gtrsim\,1$ \% would indicate new
physics~\cite{Hou:2006du}.  
These predictions provide an important standard-model reference for
verifying new physics from the $B^0\to J/\psi K_S$ data.

\begin{acknowledgments}
I am grateful to the organizers of CKM2006 for a stimulating workshop. 
I would like to thank Hsiang-nan Li for his collaboration on the
 research presented in this talk. 
This work was supported by the U.S. Department of Energy under Grant
 No. DE-FG02-90ER40542. 
\end{acknowledgments}

\end{document}